\documentclass[journal]{IEEEtran}

\usepackage[T1]{fontenc}
\usepackage{cite}
\usepackage{amsmath,amssymb,amsfonts}
\usepackage{algorithmic}
\usepackage{graphicx}
\usepackage{textcomp}
\usepackage{hyphenat}
\usepackage{hyperref}
\usepackage{siunitx}
\usepackage{booktabs} 

\newcommand{\FigOne}{
\begin{figure*}[!ht]
	\centering
	\includegraphics[width=0.96\textwidth, trim=0 0 0 0, clip]{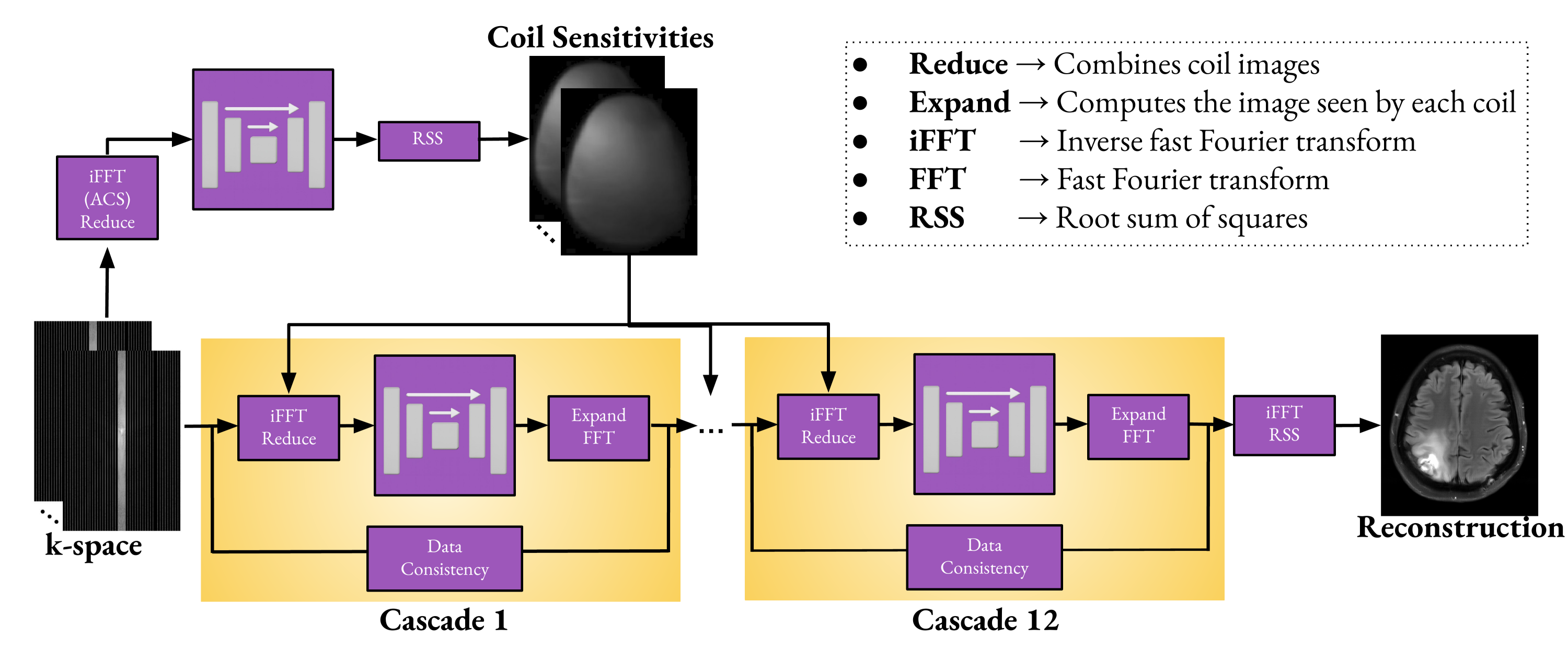}
	\caption{E2E VarNet architecture. The input is the undersampled multi\hyp coil k\hyp space and the output is the reconstructed image. In each cascade, the inverse fast Fourier transform (iFFT) is applied to each undersampled k\hyp space, and the resulting images are weighted with the corresponding coil sensitivities and are combined into a single image using the \textit{reduce operator} \cite{sriram2020end}. The combined image is processed by a U\hyp Net, whose output is then expanded \cite{sriram2020end} back into individual coil images. These coil images are transformed to k\hyp space using the FFT, after which data consistency is enforced. The final image is obtained using the root sum of squares (RSS) on the individual coil images obtained from the 12$^{\rm th}$ cascade.}\label{fig1:Architecture}
\end{figure*}}

\newcommand{\FigTwo}{
\begin{figure*}[!ht]
	\centering
	\includegraphics[width=0.96\textwidth, trim=0 0 0 0, clip]{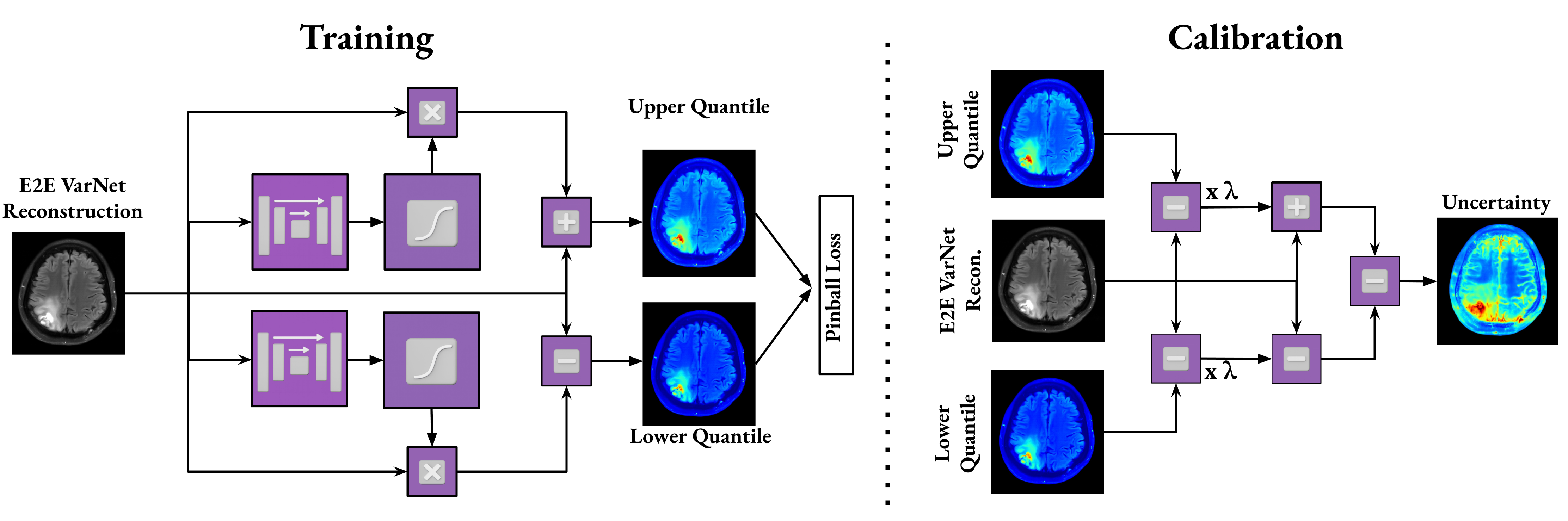}
	\caption{\textbf{(Left)} Architecture of our proposed uncertainty estimation module. The output (reconstructed MR image) of the E2E VarNet is the input of two U\hyp Nets, which learn pixelwise offsets that parameterize the lower and upper quantile bounds of the reconstruction uncertainty. Each U\hyp Net output passes through a sigmoid activation, is multiplied by the E2E VarNet reconstruction, and is either added to or subtracted from the E2E VarNet's reconstruction to compute the upper and lower bound, respectively. The output that is added is treated as the upper quantile interval and the one that is subtracted is the lower quantile interval. \textbf{(Right)} In Calibration, the predicted offsets are scaled by the calibration factor $\lambda$ which is computed using conformal prediction in the calibration set. The uncertainty map is computed by subtracting the calibrated lower from the calibrated upper bound.}\label{fig2:Architecture}
\end{figure*}}

\newcommand{\FigThree}{
\begin{figure}[!ht]
	\centering
	\includegraphics[width=0.96\columnwidth, trim=0 0 0 0, clip]{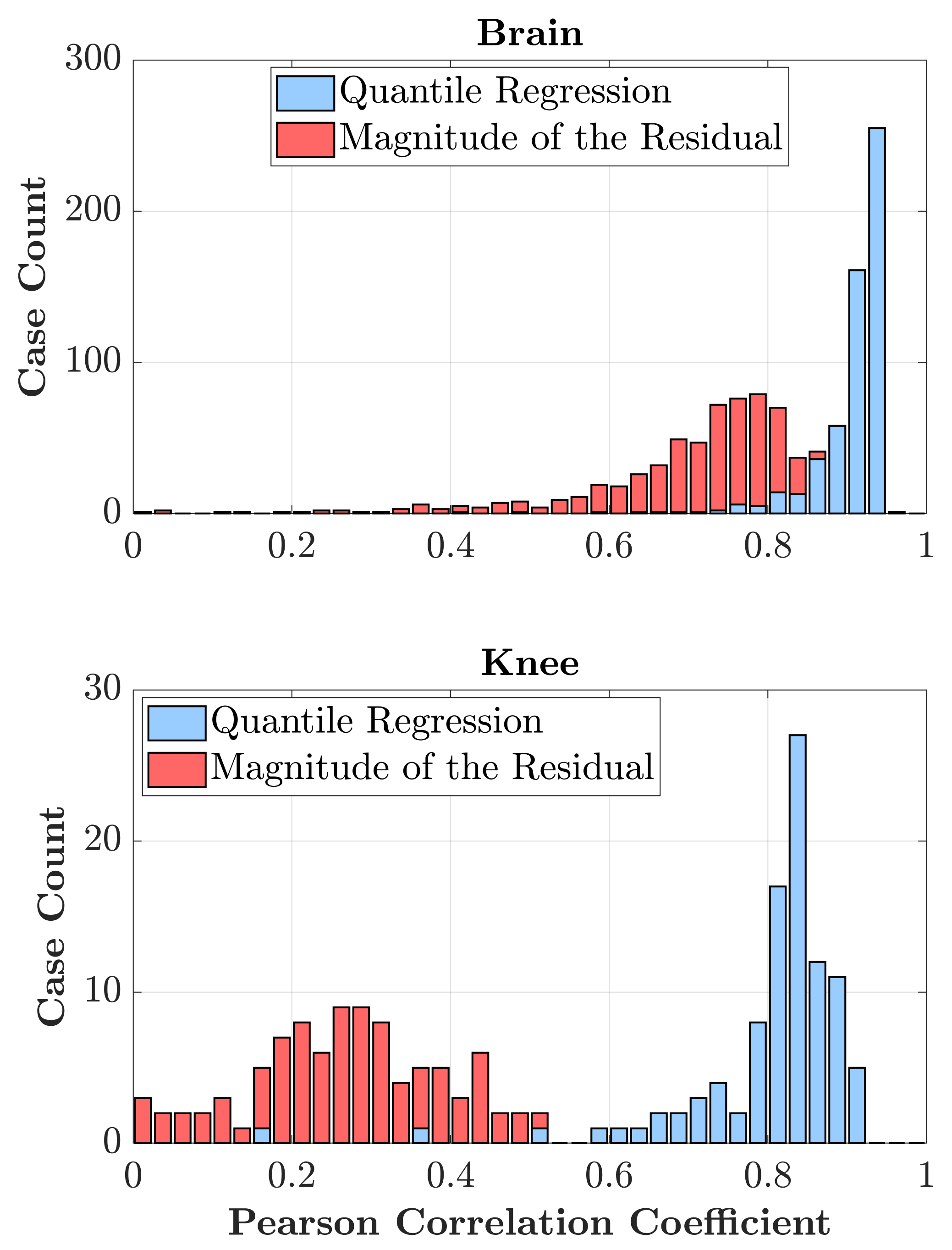}
	\caption{Distribution of Pearson correlations for the $4\times$ accelerated brain \textbf{(top)} and knee \textbf{(bottom)} reconstructions. The histograms compare QR\hyp based (blue) and ResM\hyp based (red) uncertainty estimates with the true reconstruction error for each case in the test datasets. QR\hyp based correlations are systematically higher for both anatomies, indicating improved alignment with the true reconstruction error.}\label{fig3:Result}
\end{figure}}

\newcommand{\FigFour}{
\begin{figure*}[!ht]
	\centering
	\includegraphics[width=0.96\textwidth, trim=0 0 0 0, clip]{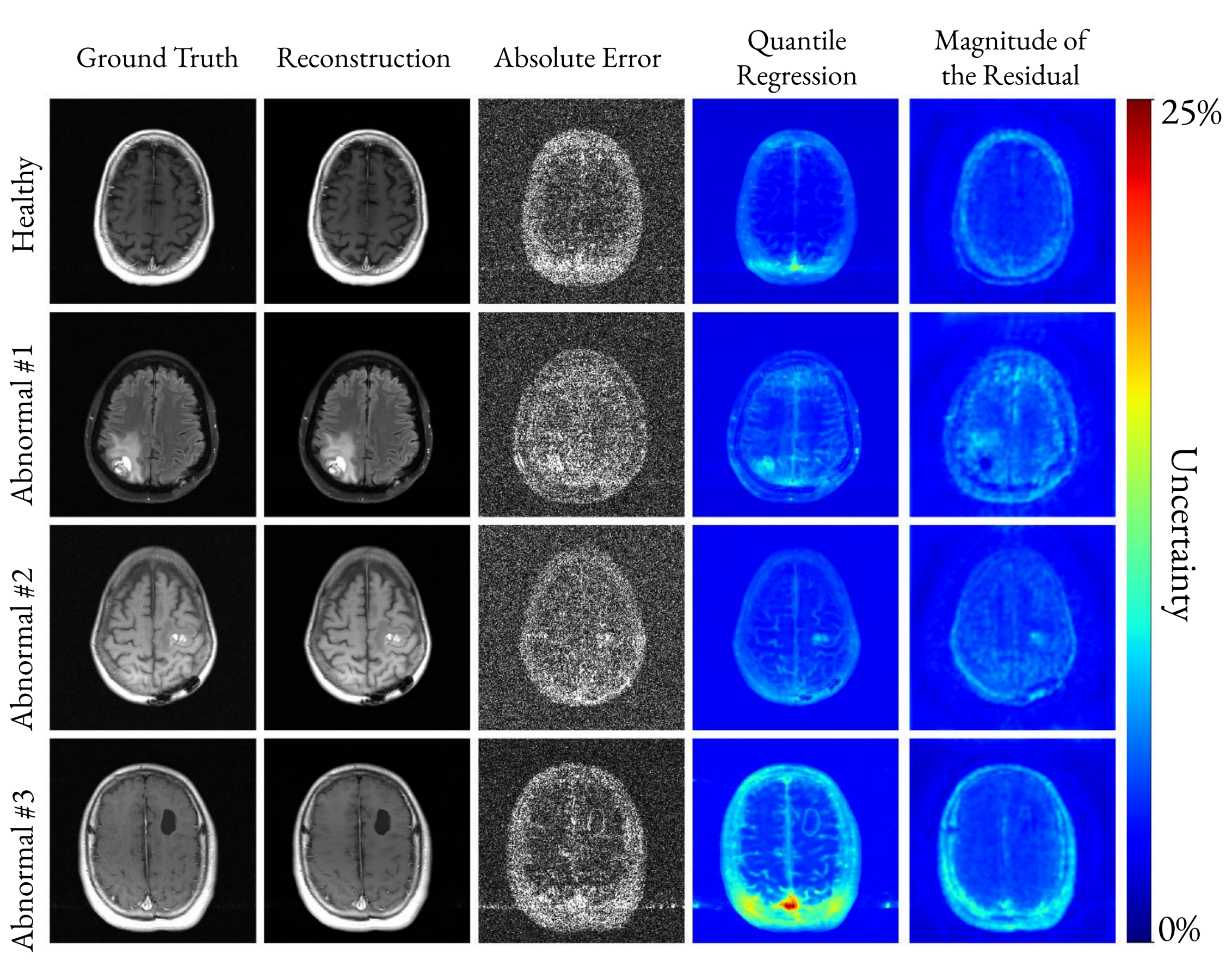}
	\caption{Comparison between the windowed absolute error (magnified 50 times), the QR\hyp based, and the ResM\hyp based uncertainty for one healthy and three abnormal brain cases. All reconstructions were performed with four\hyp fold acceleration. The QR\hyp based uncertainty closely matches the absolute error distribution and in the abnormal cases it delineates the lesions, demonstrating superior localization of uncertainty compared to ResM.}\label{fig4:Result}
\end{figure*}}

\newcommand{\FigFive}{
\begin{figure*}[!ht]
	\centering
	\includegraphics[width=0.96\textwidth, trim=0 0 0 0, clip]{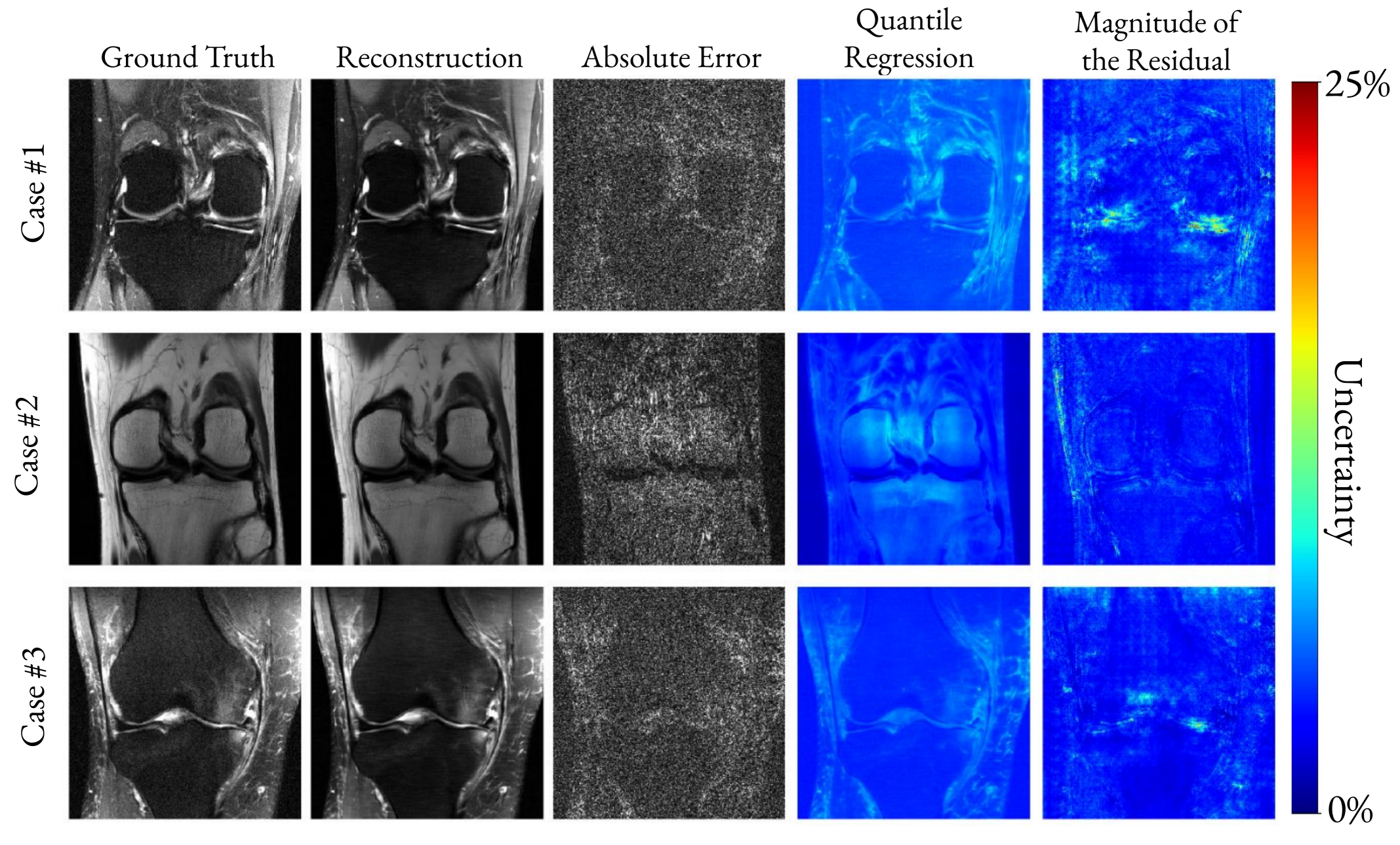}
	\caption{Comparison between the windowed absolute error (magnified 50 times), the QR\hyp based, and the ResM\hyp based uncertainty for three knees. All reconstructions were performed with four\hyp fold acceleration. The QR\hyp based uncertainty qualitatively matches the absolute error distribution unlike the ResM\hyp based approach.}\label{fig5:Result}
\end{figure*}}

\newcommand{\FigSix}{
\begin{figure*}[!ht]
	\centering
	\includegraphics[width=0.96\textwidth, trim=0 0 0 0, clip]{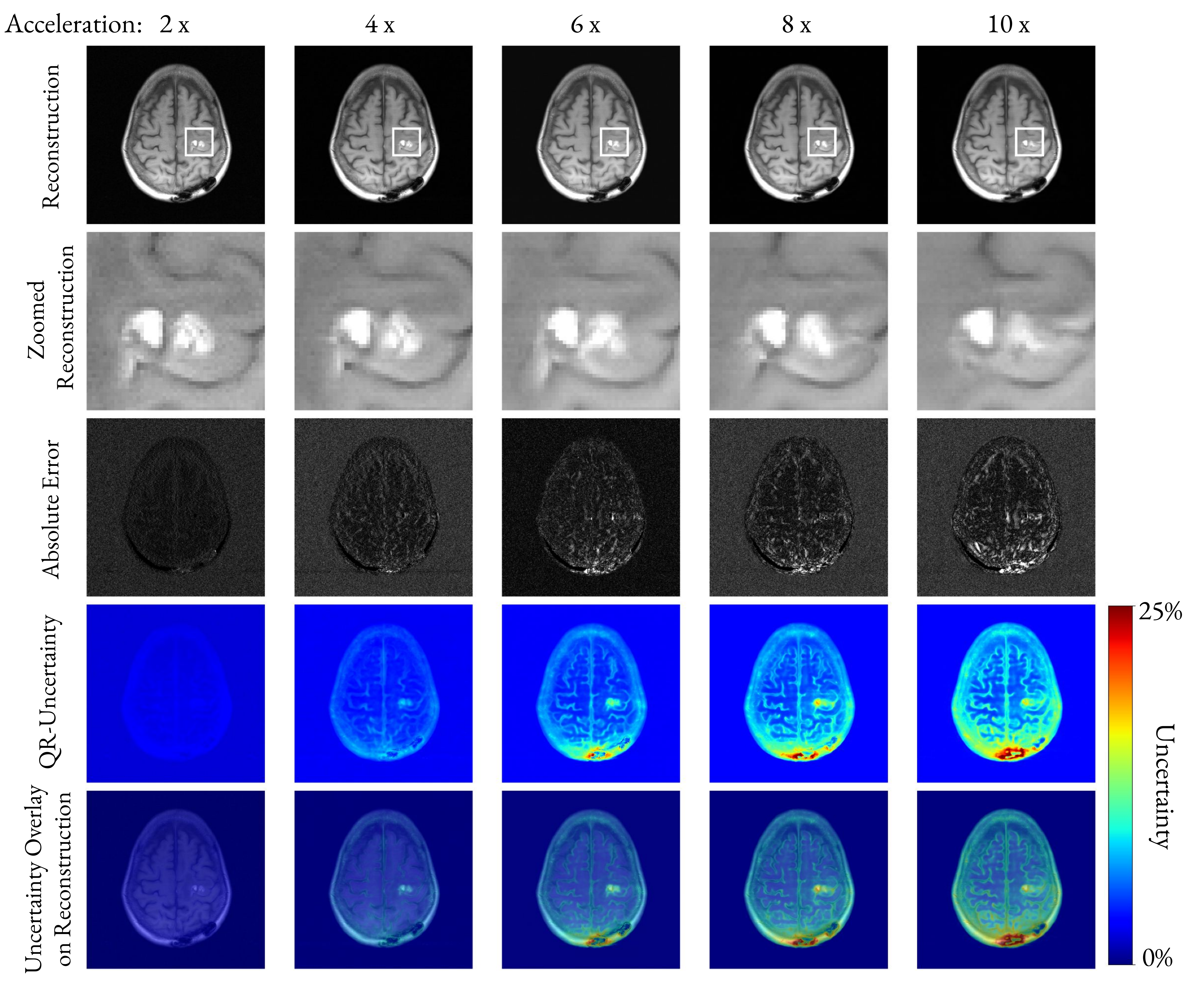}
	\caption{Reconstructed images, zoomed lesion views, absolute errors with the ground\hyp truth, and corresponding QR\hyp based uncertainty maps for one abnormal brain for five acceleration factors. Starting at four\hyp fold acceleration, the QR\hyp based uncertainty increases at the location of the lesion and reflects the posterior susceptibility artifact from a prior craniotomy. The zoomed panels facilitate visualization of the subtle lesion differences across acceleration factors. The bottom row overlays the uncertainty heatmaps on the corresponding reconstructed images to better illustrate that regions of high uncertainty spatially coincide with the lesion and the susceptibility artifact.}\label{fig6:Result}
\end{figure*}}

\newcommand{\FigSeven}{
\begin{figure*}[!ht]
	\centering
	\includegraphics[width=0.88\textwidth, trim=0 0 0 0, clip]{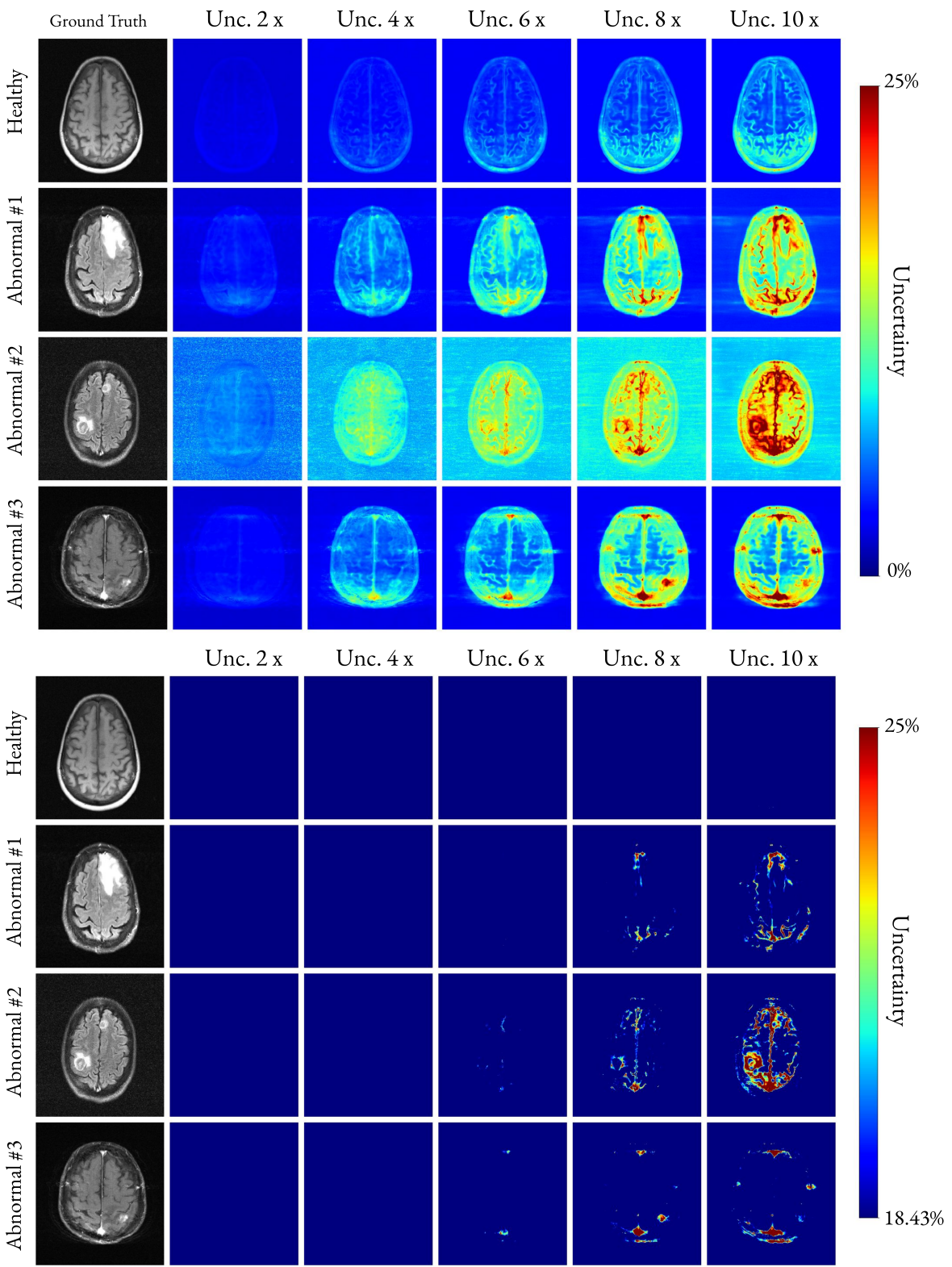}
	\caption{QR-based uncertainty maps for a healthy brain and three abnormal brain cases across increasing acceleration factors. For each case, uncertainty maps are shown for $2$, $4$, $6$, $8$, and $10\times$ acceleration to illustrate how uncertainty evolves with undersampling. \textbf{Top:} unthresholded uncertainty maps (lower bound set to zero). \textbf{Bottom:} uncertainty maps thresholded using the maximum uncertainty observed at $4\times$ acceleration. Thresholding highlights regions of elevated uncertainty at higher accelerations, corresponding to lesions and post\hyp treatment effects. Note that the background color of abnormal case 3 in the top panel, appears brighter and noisier compared the other three cases. This is expected as there is a higher level of background noise corrupting the ground\hyp truth, and the model is uncertain about its distribution.}\label{fig7:Result}
\end{figure*}}

\newcommand{\FigEight}{
\begin{figure}[!ht]
	\centering
	\includegraphics[width=0.96\columnwidth, trim=0 0 0 0, clip]{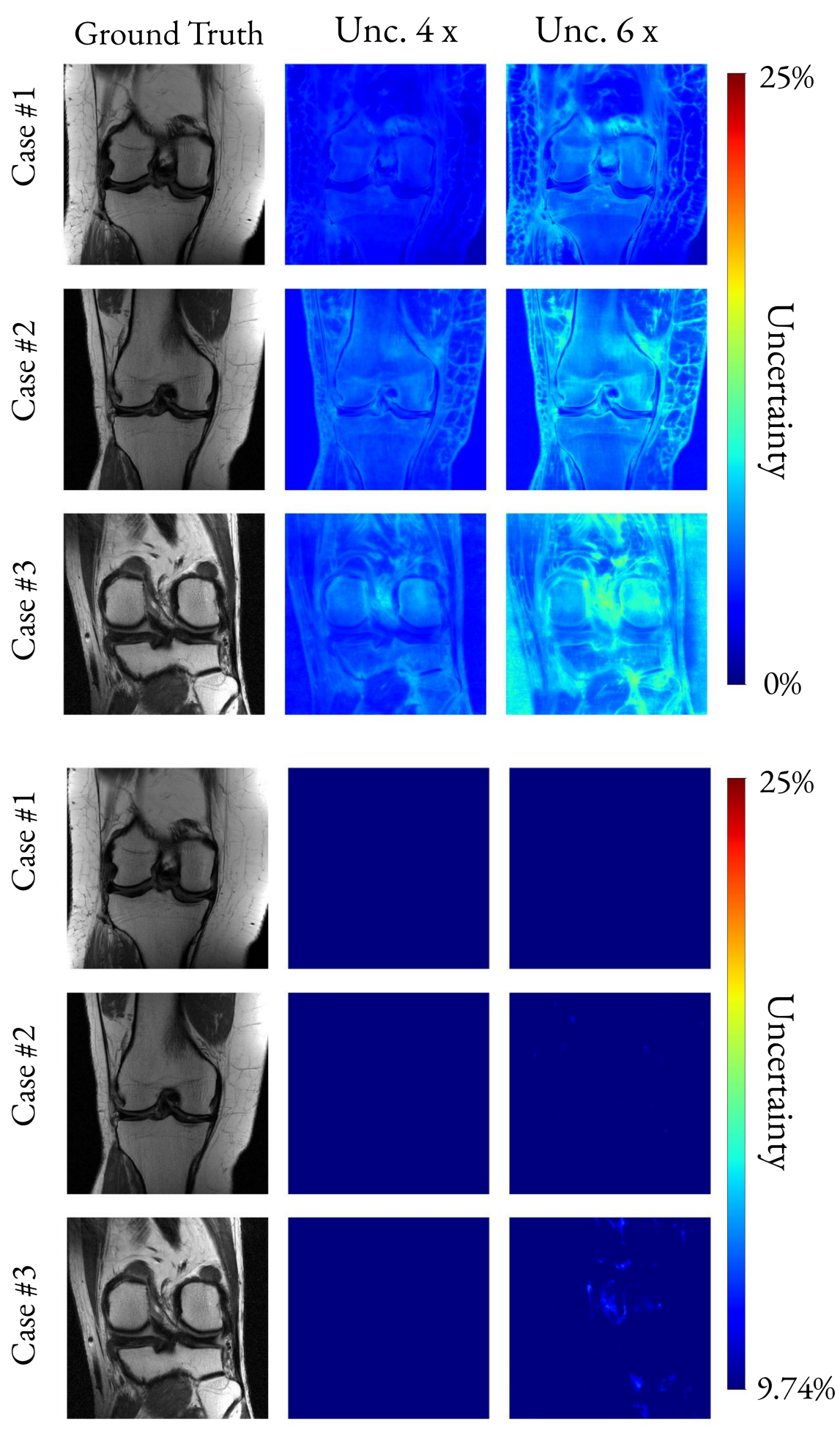}
	\caption{QR\hyp based uncertainty maps for three knee cases at increasing acceleration. For each case, uncertainty maps are shown for $4$ and $6\times$ acceleration to illustrate how uncertainty changes with undersampling in knee MRI. \textbf{Top:} unthresholded uncertainty maps (lower bound set to zero). \textbf{Bottom:} uncertainty maps thresholded using the maximum uncertainty observed at $4\times$ acceleration. Thresholding highlights regions where uncertainty exceeds levels associated with clinically reliable reconstructions.}\label{fig8:Result}
\end{figure}}

\newcommand{\TbOne}{
\begin{table*}[!ht]
\fontsize{8.5}{12}\selectfont   
\caption{Mean and standard deviation of SSIM and uncertainty} 
\label{tb:1}
\begin{tabular}{p{1.5cm}|
    >{\centering\arraybackslash}p{0.6cm}|
    >{\centering\arraybackslash}p{1.3cm}|
    >{\centering\arraybackslash}p{0.6cm}|
    >{\centering\arraybackslash}p{1.3cm}|
    >{\centering\arraybackslash}p{1.3cm}|
    >{\centering\arraybackslash}p{0.6cm}|
    >{\centering\arraybackslash}p{1.3cm}|
    >{\centering\arraybackslash}p{0.6cm}|
    >{\centering\arraybackslash}p{1.3cm}|
    >{\centering\arraybackslash}p{0.6cm}|
    >{\centering\arraybackslash}p{1.3cm}}
    
    \toprule
    
    \textbf{Acceleration}                   &
    \multicolumn{2}{c|}{\textbf{2$\times$}} &
    \multicolumn{3}{c|}{\textbf{4$\times$}} &
    \multicolumn{2}{c|}{\textbf{6$\times$}} &
    \multicolumn{2}{c|}{\textbf{8$\times$}} &
    \multicolumn{2}{c}{\textbf{10$\times$}} \\

    \cmidrule(lr){1-12}

    \textbf{Metric}    & 
    \textbf{SSIM}      & 
    \textbf{QR} (\%)   & 
    \textbf{SSIM}      & 
    \textbf{QR (\%)}   & 
    \textbf{ResM (\%)} & 
    \textbf{SSIM}      & 
    \textbf{QR} (\%)   & 
    \textbf{SSIM}      & 
    \textbf{QR} (\%)   & 
    \textbf{SSIM}      & 
    \textbf{QR} (\%)   \\

    \midrule
    
    \textbf{T1}     
        & 0.979 
        & 2.1 $\pm$ 0.82 
        & 0.966
        & 3.4 $\pm$ 1.85 
        & 3.8 $\pm$ 1.82
        & 0.960  
        & 4.0 $\pm$ 2.54 
        & 0.954  
        & 4.8 $\pm$ 3.61 
        & 0.950  
        & 5.4 $\pm$ 4.29 \\
    \textbf{T1PRE}  
        & 0.971
        & 2.6 $\pm$ 1.00   
        & 0.956
        & 4.0 $\pm$ 1.99 
        & 4.3 $\pm$ 2.36   
        & 0.949
        & 4.7 $\pm$ 2.73   
        & 0.942
        & 5.6 $\pm$ 3.88   
        & 0.937
        & 6.4 $\pm$ 4.74 \\
    \textbf{T1POST}   
        & 0.980
        & 2.2 $\pm$ 0.87   
        & 0.965
        & 3.7 $\pm$ 2.18 
        & 3.3 $\pm$ 1.94   
        & 0.957
        & 4.4 $\pm$ 3.06   
        & 0.950
        & 5.4 $\pm$ 4.35   
        & 0.944
        & 6.2 $\pm$ 5.34 \\
    \textbf{T2}       
        & 0.974
        & 2.6 $\pm$ 1.65   
        & 0.959
        & 4.1 $\pm$ 2.84  
        & 4.0 $\pm$ 2.71   
        & 0.951 
        & 5.0 $\pm$ 3.67  
        & 0.945 
        & 6.1 $\pm$ 5.10 
        & 0.940
        & 7.0 $\pm$ 6.17 \\
    \textbf{FLAIR}  
        & 0.957
        & 3.4 $\pm$ 1.46 
        & 0.931 
        & 5.3 $\pm$ 2.75 
        & 6.2 $\pm$ 3.55 
        & 0.919
        & 6.1 $\pm$ 3.58 
        & 0.909
        & 7.3 $\pm$ 4.84 
        & 0.902
        & 8.3 $\pm$ 5.76 \\
    \textbf{Knee}  
        & -
        & - 
        & 0.927
        & 4.6 $\pm$ 2.15 
        & 3.9 $\pm$ 2.18 
        & 0.911
        & 6.3 $\pm$ 3.05 
        & -
        & - 
        & -
        & - \\
    
    \bottomrule
\end{tabular}
\end{table*}}

\newcommand{\TbTwo}{
\begin{table*}[!ht]
\caption{Correlation between uncertainty and reconstruction error}
\label{tb:2}
\begin{tabular}{p{3cm}|
    >{\centering\arraybackslash}p{1.85cm}|
    >{\centering\arraybackslash}p{1.85cm}|
    >{\centering\arraybackslash}p{1.85cm}|
    >{\centering\arraybackslash}p{1.85cm}|
    >{\centering\arraybackslash}p{1.85cm}|
    >{\centering\arraybackslash}p{1.85cm}}
    
    \toprule
    
    \textbf{Acceleration}                   &
    \textbf{2$\times$}                      &
    \multicolumn{2}{c|}{\textbf{4$\times$}} &
    \textbf{6$\times$}                      &
    \textbf{8$\times$}                      &
    \textbf{10$\times$}                    \\

    \cmidrule(lr){1-7}

    \textbf{Metric} (\%) & 
    \textbf{QR}          & 
    \textbf{QR}          &
    \textbf{ResM}        &  
    \textbf{QR}          & 
    \textbf{QR}          & 
    \textbf{QR}         \\
    
    \cmidrule(lr){1-7}
    
    \multicolumn{7}{c}{\textbf{Brain}} \\

    \midrule
    
    \textbf{Pearson}              
        & 0.82 $\pm$ 0.12 
        & 0.91 $\pm$ 0.10 
        & 0.69 $\pm$ 0.22 
        & 0.92 $\pm$ 0.09
        & 0.91 $\pm$ 0.09
        & 0.90 $\pm$ 0.09 \\
    
    \textbf{Region-Pearson}  
        & 0.92 $\pm$ 0.12
        & 0.96 $\pm$ 0.11
        & 0.73 $\pm$ 0.27
        & 0.97 $\pm$ 0.09
        & 0.97 $\pm$ 0.10
        & 0.97 $\pm$ 0.09 \\
    
    \textbf{Spearman}              
        & 0.70 $\pm$ 0.18 
        & 0.76 $\pm$ 0.18 
        & 0.63 $\pm$ 0.22 
        & 0.76 $\pm$ 0.18
        & 0.77 $\pm$ 0.18
        & 0.77 $\pm$ 0.19 \\
    
    \textbf{Region-Spearman} 
        & 0.75 $\pm$ 0.20 
        & 0.79 $\pm$ 0.22 
        & 0.54 $\pm$ 0.31 
        & 0.84 $\pm$ 0.20
        & 0.86 $\pm$ 0.19
        & 0.87 $\pm$ 0.18 \\
    
    \cmidrule(lr){1-7}
    
    \multicolumn{7}{c}{\textbf{Knee}} \\

    \midrule
    
    \textbf{Pearson}              
        & - 
        & 0.80 $\pm$ 0.16
        & 0.24 $\pm$ 0.27 
        & 0.82 $\pm$ 0.15
        & -
        & - \\
    
    \textbf{Region-Pearson}        
        & - 
        & 0.89 $\pm$ 0.17
        & 0.16 $\pm$ 0.36 
        & 0.91 $\pm$ 0.17
        & -
        & - \\
    
    \textbf{Spearman}                
        & - 
        & 0.72 $\pm$ 0.20
        & 0.26 $\pm$ 0.25 
        & 0.75 $\pm$ 0.20
        & -
        & - \\
    
    \textbf{Region-Spearman}       
        & - 
        & 0.79 $\pm$ 0.23
        & 0.12 $\pm$ 0.34 
        & 0.81 $\pm$ 0.23
        & -
        & - \\
        
    \bottomrule
\end{tabular}
\end{table*}}

\begin{document}
\bstctlcite{IEEEexample:BSTcontrol}

\title{Pixelwise Uncertainty Quantification of Accelerated MRI Reconstruction}
\author{Ilias I. Giannakopoulos$^\dagger$, \IEEEmembership{Senior Member, IEEE}, Lokesh B Gautham Muthukumar$^\dagger$, Yvonne W. Lui, and Riccardo Lattanzi, \IEEEmembership{Senior Member, IEEE}
\thanks{$^\dagger$ indicates equivalent contribution. This work was supported in part by NIH K99 EB035163, in part by NIH R01 EB024536, and was performed under the Rubric of the Center for Advanced Imaging Innovation and Research (CAI$^2$R, www.cai2r.net), an NIBIB National Center for Biomedical Imaging and Bioengineering (NIH P41 EB017183).}
\thanks{Ilias I. Giannakopoulos, Lokesh B Gautham Muthukumar, Yvonne W. Lui, and Riccardo Lattanzi are with the Bernard and Irene Schwartz Center for Biomedical Imaging, Department of Radiology, NYU Grossman School of Medicine, 10016, New York, NY, United States of America (e-mail:ilias.giannakopoulos@nyulangone.org, LokeshGautham.BoominathanMuthukumar@nyulangone.org, yvonne.lui@nyulangone.org, riccardo.lattanzi@nyulangone.org).}
\thanks{Lokesh B Gautham Muthukumar is also with the Courant Institute of Mathematical Sciences, NYU, 10012, New York, NY, United States of America.}
\thanks{Yvonne W. Lui and Riccardo Lattanzi are also with the Center for Advanced Imaging Innovation and Research (CAI$^2$R), Department of Radiology, NYU Grossman School of Medicine, 10016, New York, NY, United States of America.}}

\maketitle

\begin{abstract}
The goal of this work is to introduce an automated method to assess the quality of under-sampled MRI reconstructions. We propose a general framework for pixel-wise uncertainty quantification in accelerated MRI reconstructions, enabling automatic identification of unreliable regions without using ground-truth fully-sampled reference images. Our method integrates conformal quantile regression with learning\hyp based image reconstruction methods to estimate statistically rigorous pixel-wise uncertainty intervals. We trained and evaluated our model on Cartesian undersampled brain and knee data obtained from the fastMRI dataset using acceleration factors ranging from 2 to 10. An end-to-end Variational Network was used for image reconstruction. Quantitative experiments demonstrate strong agreement between predicted uncertainty maps and true reconstruction error. Using our method, the corresponding Pearson correlation coefficient was higher than 90\% at acceleration levels at and above four-fold; whereas it dropped to less than 70\% when the uncertainty was computed using a simpler a heuristic notion (magnitude of the residual). Qualitative examples further show the uncertainty maps based on quantile regression capture the magnitude and spatial distribution of reconstruction errors across acceleration factors, with regions of elevated uncertainty aligning with pathologies and artifacts. The proposed framework enables evaluation of reconstruction quality without access to fully-sampled ground-truth reference images. It represents a step toward adaptive MRI acquisition protocols that may be able to dynamically balance scan time and diagnostic reliability.
\end{abstract}


\FigOne

\section{Introduction}
\label{sec:introduction}
\IEEEPARstart{M}{agnetic} resonance imaging (MRI) provides unparalleled soft\hyp tissue contrast and plays a central role in modern disease diagnostics. However, its high diagnostic value is counterbalanced by inherently long acquisition times, which limit patient throughput, increase susceptibility to motion artifacts, reduce patient comfort, and systemically, raise operational costs and decrease accessibility. Rapid MRI techniques, such as parallel imaging (PI) \cite{sodickson1997simultaneous, pruessmann1999sense, griswold2002generalized} and compressed sensing (CS) \cite{lustig2007sparse, uecker2008image},can mitigate these limitations by reconstructing images from undersampled k\hyp space data. More recently, deep learning–based unrolled reconstruction networks achieved superior reconstruction quality compared with traditional methods \cite{hammernik2018learning, aggarwal2018modl, eo2018kiki, sriram2020end, arvinte2021deep, jun2021joint, giannakopoulos2024accelerated}. Clinical evaluation studies demonstrated that these networks produce diagnostically interchangeable reconstructions equivalent to fully sampled references at four\hyp fold acceleration \cite{recht2020using, lin2021artificial, johnson2023deep, giannakopoulos2024accelerated}, which led to their recent U.S. Food and Drug Administration (FDA) clearance. Diagnostic quality can be maintained even at higher accelerations in certain cases. For example, the study in \cite{radmanesh2022exploring} reported that 100\% of 4$\times$, 97\% of 6$\times$, 62\% of 8$\times$, 17\% of 10$\times$, and 3\% of 12$\times$ accelerated T2\hyp weighted images reconstructed using the end\hyp to\hyp end variational network (E2E VarNet) \cite{sriram2020end} are appropriate for general\hyp purpose diagnostic imaging. For healthy subjects, 10$\times$ acceleration factor can result in acceptable images, as these networks are primarily trained on cases without pathologies \cite{johnson2025prostate}. This suggests that some fraction of MRI scans performed today could, in principle, be done faster than what current conservative, fixed\hyp acceleration protocols allow.
\par
Despite the potential for faster scans, high acceleration levels are not routinely implemented in clinical MRI protocols. The primary limiting factor is the absence of an automatic mechanism to assess the reliability of reconstructed images when no fully sampled reference is available. Without a quantitative measure of confidence, increasing the acceleration factor beyond established values is risky, as undersampling may lead to network hallucinations \cite{bhadra2021hallucinations} or pathological feature suppression \cite{kiryu2023clinical}, ultimately compromising diagnostic accuracy. Consequently, current clinical practice favors conservative, fixed acceleration factors, typically set to two for non\hyp learning–based reconstructions \cite{pruessmann1999sense} and four for learning\hyp based reconstructions \cite{sriram2020end}. To enable higher accelerations while preserving diagnostic validity, there is a critical need for methods capable of estimating the reconstruction error directly from undersampled data. Uncertainty quantification \cite{kendall2017uncertainties, abdar2021review, angelopoulos2022image} provides a principled framework toward that goal by estimating confidence intervals for reconstructed images and effectively producing an ``error map'' when the ground\hyp truth reference image is unavailable.
\par
Early attempts to quantify uncertainty in MRI reconstructions were based on probabilistic formulations of deep reconstruction networks. For example, \cite{edupuganti2020uncertainty} proposed a variational autoencoder framework that learns to predict pixel\hyp wise residual magnitudes for a pre\hyp trained reconstruction network. The resulting uncertainty maps provide a useful visualization of reconstruction reliability, but the approach remains heuristic and lacks statistical guarantees on the predicted uncertainty levels. In \cite{tanno2019uncertainty}, a Bayesian heteroscedastic uncertainty framework is used to highlight regions of potential reconstruction failure, but does not yield calibrated or distribution\hyp free confidence intervals. The work in \cite{narnhofer2021bayesian} introduced a Bayesian variational formulation of the learned variational network, in which the parameters of the regularizer are modeled as random variables drawn from a learned multivariate Gaussian distribution. Sampling these parameters enables estimation of pixel\hyp wise variance maps that reflect regions of high uncertainty. While this framework offers a theoretically grounded means of visualizing uncertainty, it incurs substantial computational overhead and does not ensure calibrated or interpretable confidence intervals. PixCUE \cite{ekanayake2025pixcue} reformulated MRI reconstruction as a pixel\hyp classification problem, where uncertainty is derived from the variance of the softmax output distribution. While this approach enables efficient joint estimation of reconstruction and uncertainty, it remains heuristic and lacks statistical calibration or formal coverage guarantees. 
\par
A major advance toward statistically valid uncertainty estimation came with the work of \cite{angelopoulos2022image}, who introduced a distribution\hyp free framework for image\hyp to\hyp image regression based on conformal prediction. The proposed methods produce pixel\hyp wise uncertainty intervals with finite\hyp sample coverage guarantees by calibrating heuristic uncertainty estimates on a held\hyp out calibration dataset. \cite{wang2022rigorous} applied this approach to MRI reconstruction using the magnitude of the residual (ResM) as an uncertainty estimator. Despite encouraging results, ResM yields symmetric confidence bounds that limit interpretability and can inflate the estimated uncertainty range. Moreover, the residuals learned during training are often smaller than those encountered during testing, leading to unreliable uncertainty estimates that correlate poorly with the true reconstruction error \cite{angelopoulos2022image}.
\par
We demonstrate a statistically rigorous uncertainty estimation framework based on conformalized quantile regression (QR) \cite{angelopoulos2022image} {\color{black}that can be retrained to work with any learning\hyp based image reconstruction method}. The proposed module directly estimates the conditional quantiles of the reconstructed image distribution, thereby producing statistically valid and spatially resolved uncertainty intervals. A similar QR\hyp based strategy has recently been applied in CUTE\hyp MRI \cite{fischer2025cutemri}, where conformal calibration is used to control uncertainty in downstream clinical metrics to enable time\hyp adaptive acquisitions. In contrast, our work focuses on pixel\hyp wise uncertainty estimation of the reconstructed image itself, providing spatially localized confidence maps. A preliminary version of this work was presented at the 2026 Annual Meeting of the International Society for Magnetic Resonance in Medicine \cite{giannakopoulos2026pixelwise, giannakopoulos2026trainable}.

\FigTwo
\TbOne

\section{Methods}

\subsection{Dataset}

Data used in this work were obtained from the NYU fastMRI Initiative database (\href{fastmri.med.nyu.edu}{fastmri.med.nyu.edu}) \cite{zbontar2018fastmri, knoll2020fastmri}. We used both the brain and knee fastMRI datasets. For brain experiments, the networks were trained on the fastMRI brain training dataset (4469 volumes). The fastMRI brain validation dataset (1378 volumes) was randomly divided into two equal subsets for validation and calibration (689 volumes each; see Section~\ref{sec:cal}). The full brain test dataset (558 volumes) was used for testing. For the knee experiments, networks were trained on the fastMRI knee training dataset (973 volumes). Since the ground\hyp truth reference images are not available in the fastMRI knee test dataset, we divided the knee validation dataset into 49 volumes for validation, 50 for calibration, and 99 for testing.

\subsection{Undersampling Strategy}
Cartesian undersampling was used for all experiments. For the brain dataset, we retained 16\%, 8\%, 5.3\%, 4\%, or 3\% of the central k\hyp space as autocalibration signal (ACS) lines and uniformly sampled the remaining k\hyp space to achieve acceleration factors of 2, 4, 6, 8, and 10, respectively. All models were trained and tested using fixed undersampling masks, resulting in five separate models for reconstruction. For the knee dataset, 8\% or 5.3\% of the central k\hyp space was retained as ACS lines, and the remaining k\hyp space was uniformly sampled to achieve an acceleration factor of 4 or 6, respectively, resulting in two separate models for reconstruction. 

\subsection{Baseline Reconstruction Network}

\TbTwo
\FigThree

We used the E2E VarNet as our image reconstruction method \cite{sriram2020end}. The model's architecture is shown in the top panel of Figure~\ref{fig1:Architecture}. We used a total of 12 cascades. In each cascade, the U\hyp Net \cite{ronneberger2015u} used 32 feature channels, 4 layers of average pooling and transpose convolutions, each with a kernel size, stride, and padding of 2, 2, and 0, respectively. The convolution layers used a kernel size of 3 with both padding and stride set to 1. Leaky ReLU activation functions were used with a negative slope of 0.2. The U\hyp Net for the coil sensitivity estimation had the same parameters except the number of channels which was reduced to $8$. The input tensors to all U\hyp Nets were normalized to ensure that each channel had a mean of 0 and a standard deviation of 1. We used a batch size of 1 and trained on a high\hyp performance computing cluster using four NVIDIA A100 Tensor Core GPUs, each with \SI{80}{\giga\byte} of memory. Our loss function was the structural similarity index measure (SSIM) \cite{wang2003multiscale}. Training was performed using the AdamW optimizer with a learning rate of 0.0003 for 210{,}000 iteration steps. A warm\hyp up ramp was applied during the first 7{,}500 steps, followed by cosine annealing after 140{,}000 steps. The overall number of trainable parameters was 93.6 million and the training times was approximately two and a half days for all models.

\subsection{Uncertainty Estimation Framework}

\subsubsection{Uncertainty Module Architecture}

Our uncertainty estimation module architecture is shown in Figure~\ref{fig2:Architecture} (left). The output reconstruction of the E2E VarNet is used as the input to this module. The architecture consists of two U\hyp Nets with the same settings as those used in the E2E VarNet cascades, except that the number of input and output channels is set to 1 (real data) instead of 2 (complex data). Each U\hyp Net output passes through a sigmoid activation to scale the values between 0 and 1, after which it is multiplied by the E2E VarNet reconstruction to match the same intensity range as the reconstructed image. The resulting outputs are used as the upper and lower quantile intervals, respectively. When symmetric upper and lower quantiles are desired, a single U\hyp Net can be used. 

\subsubsection{Pixelwise Quantile Regression Training}

The two U\hyp Nets described above directly parameterize the pixelwise lower and upper quantile functions $\tilde{l}(\mathbf{x})$ and $\tilde{u}(\mathbf{x})$, and are trained using QR with the quantile loss. In pixelwise QR \cite{koenker1978regression, romano2019conformalized}, assuming we target an $\alpha$\hyp level uncertainty interval of the E2E VarNet reconstructed image ($\mathbf{x}$), the upper quantile $\tilde{u}(\mathbf{x})$ is trained to estimate the $(1-\alpha/2)$ conditional quantile, and the lower quantile $\tilde{l}(\mathbf{x})$ to estimate the $(\alpha/2)$ conditional quantile. These pixelwise quantiles can be estimated using the quantile (pinball) loss \cite{angelopoulos2022image}, defined as
\begin{equation}
    \mathcal{L}_{\alpha}(\hat{q}_{\alpha}(\mathbf{x}), \mathbf{y}) =
    \begin{cases}
        \alpha \, (\mathbf{y} - \hat{q}_{\alpha}(\mathbf{x})), & \text{if } \mathbf{y} > \hat{q}_{\alpha}(\mathbf{x}), \\[6pt]
        (1 - \alpha) \, (\hat{q}_{\alpha}(\mathbf{x}) - \mathbf{y}), & \text{otherwise.}
    \end{cases}
\end{equation}
Here, $\mathbf{y}$ denotes the ground\hyp truth reference image and $\hat{q}_{\alpha}(\mathbf{x})$ denotes the predicted $\alpha$\hyp quantile of $\mathbf{y}$ conditioned on $\mathbf{x}$. Because the upper and lower quantiles correspond to different conditional quantiles, they are trained using separate losses. The total loss is therefore written as
\begin{equation}
\mathcal{L}(\mathbf{x}, \mathbf{y}) = \mathcal{L}_{\alpha/2}(\tilde{l}(\mathbf{x}), \mathbf{y}) + \mathcal{L}_{1 - \alpha/2}(\tilde{u}(\mathbf{x}), \mathbf{y}).
\end{equation}
After training, the lower and upper quantile estimates are expected to converge asymptotically \cite{koenker1978regression, chaudhuri1991global, zhou1996direct, zhou1998statistical, takeuchi2006nonparametric, steinwart2011estimating} to the true conditional $(\alpha/2)$\hyp and $(1-\alpha/2)$\hyp quantiles, respectively.
\par
Training is performed in a supervised manner using $(\mathbf{x}, \mathbf{y})$ pairs. The uncertainty module is trained separately from the E2E VarNet, using the same optimizer, learning rate, scheduler, and hyperparameters. The E2E VarNet is kept frozen. During reconstruction, one can either use the E2E VarNet to generate reconstructions on the fly, or the reconstructions can be precomputed for all training and validation data and loaded from memory. We set $\alpha = 90\%$ for all experiments. The number of trainable parameters was 15.5 million and the training time was approximately one day for all models.

\subsubsection{Magnitude of the Residual Training}

For the ResM\hyp based approach, we assumed that $\tilde{u}(\mathbf{x})$ = $\tilde{l}(\mathbf{x})$. The loss function used in the case is written as:
\begin{equation}
    \mathcal{L}(\mathbf{x},\mathbf{y}) = \left( \tilde{u}(\mathbf{x}) - |x - y| \right)^2.
\end{equation}
The number of trainable parameters in the ResM case was 7.75 million and the training time was approximately one day for both anatomies.

\subsubsection{Conformal Prediction Calibration}\label{sec:cal}

The quantile estimates $\tilde{l}$ and $\tilde{u}$ must be calibrated to be statistically valid, meaning that the resulting uncertainty intervals should contain at least a fraction $(1-\alpha)$ of the ground\hyp truth pixel values with probability not smaller than $(1-\alpha)$. In practice, neural networks trained to predict quantiles may produce intervals that are miscalibrated, either too narrow or too wide. To guarantee valid coverage, we apply conformal prediction calibration \cite{bates2021distribution, angelopoulos2022image}, as in Figure~\ref{fig2:Architecture} (right).

In particular, we rescaled the predicted offsets between the reconstructed image $\mathbf{x}$ and the learned quantiles using a data\hyp driven correction factor $\lambda$. In standard conformal QR \cite{angelopoulos2022image}, the quantiles themselves are rescaled. However, because our architecture \ref{fig2:Architecture} scales the U-Net outputs with the images, the natural quantities to calibrate are the offsets $(\mathbf{x} - \tilde{l})$ and $(\tilde{u} - \mathbf{x})$. To perform the calibration, we start by defining the pixelwise uncertainty interval for pixel $(m,n)$ of $\mathbf{x}$ as:
\begin{align}
&\mathcal{T}_{\lambda}(\mathbf{x}_{(m,n)}) = \left[\tilde{l}_b, \tilde{u}_b\right] \notag \\
& \tilde{l}_b = \mathbf{x}_{(m,n)} - \lambda \cdot \left( \mathbf{x}_{(m,n)} - \tilde{l}(\mathbf{x}_{(m,n)}) \right) \notag \\
& \tilde{u}_b =\mathbf{x}_{(m,n)} + \lambda \cdot \left( \tilde{u}(\mathbf{x}_{(m,n)}) - \mathbf{x}_{(m,n)} \right).
\end{align}
Then, we seek the smallest $\lambda$ such that $\mathcal{T}_{\lambda}$ achieves the desired coverage. This is computed using a held\hyp out calibration dataset of $I$ image pairs $\{(\mathbf{x}^i, \mathbf{y}^i)\}_{i=1}^{I}$, where $\mathbf{x}^i$ are reconstructions from the E2E VarNet and $\mathbf{y}^i$ are the corresponding ground\hyp truth images. The value of $\lambda$ is chosen such that the average fraction of all pixels $(m,n)$ across the calibration set whose ground\hyp truth intensities fall outside their corresponding intervals $\mathcal{T}_{\lambda}(\mathbf{x}^i_{(m,n)})$ does not exceed $\alpha$. Since this fraction is estimated from a finite calibration set, a conservative upper bound $\hat{R}^{+}(\lambda)$ is used following Hoeffding's inequality \cite{hoeffding1963probability}. The final calibrated scaling factor is therefore given by
\begin{equation}
    \min \Big\{
    \lambda : \hat{R}^{+}(\lambda) \le \alpha
    \Big\}.
\end{equation}
The calibration required roughly 6 hours for each brain model and 3 hours for each knee model. {\color{black}Once computed, $\lambda$ is fixed for the corresponding trained model and is directly used during inference}. The values of $\lambda$ for the brain and QR\hyp based uncertainty were $1.31$, $1.54$, $1.59$, $1.74$, and $1.87$ for $2\times$, $4\times$, $6\times$, $8\times$, and $10\times$ acceleration, respectively. For the ResM\hyp based uncertainty, the scaling factor at $4\times$ acceleration was $1.02$. For the knee, the scaling factors for the QR\hyp based uncertainty were $1.42$ and $1.68$ for $4\times$ and $6\times$ acceleration, respectively. For the ResM\hyp based uncertainty, the scaling factor at $4\times$ acceleration was $0.82$. 

\subsection{Evaluation of Uncertainty Reliability}

To assess the correspondence between the predicted uncertainty and the true reconstruction error, we define the uncertainty map $\tilde{q}$ and the absolute error map $\tilde{e}$ as
\begin{equation}
\tilde{q} = \tilde{u}_b(\mathbf{x}) - \tilde{l}_b(\mathbf{x}), \qquad
\tilde{e} = |\mathbf{x} - \mathbf{y}|,
\end{equation}
The map $\tilde{q}$ quantifies the pixelwise uncertainty width, while $\tilde{e}$ represents the magnitude of the absolute (true) reconstruction error. Notice that $\tilde{q}$ does not have any knowledge of the ground\hyp truth $\mathbf{y}$.

\subsubsection{Correlation Metrics}

The correspondence between the predicted uncertainty $\tilde{q}$ and the true error $\tilde{e}$ is quantified using the Pearson \cite{pearson1895vii} and Spearman \cite{spearman1961general} correlation coefficients. The Pearson correlation coefficient measures the linear association between $\tilde{q}$ and $\tilde{e}$ and is defined as
\begin{equation}
r = \frac{\mathrm{cov}(\tilde{q}, \tilde{e})}
{\sigma_{\tilde{q}} \sigma_{\tilde{e}}},
\end{equation}
where $\mathrm{cov}$ denotes the covariance.

The Spearman rank correlation coefficient quantifies whether pixels with higher predicted uncertainty also tend to have higher reconstruction errors, independent of their exact numerical scaling. It is computed by applying the Pearson correlation to the rank-ordered variables, where each pixel value in $\tilde{q}$ and $\tilde{e}$ is replaced by its position in the sorted list of all pixel values within the image, yielding $R_{\tilde{q}}$ and $R_{\tilde{e}}$, respectively. The coefficient is then computed as
\begin{equation}
\rho = \frac{\mathrm{cov}(R_{\tilde{q}}, R_{\tilde{e}})}
{\sigma_{R_{\tilde{q}}} \sigma_{R_{\tilde{e}}}}.
\end{equation}
Here, $R_{\tilde{q}}$ and $R_{\tilde{e}}$ denote the rank-ordered maps, in which lower pixel values receive smaller ranks and higher pixel values receive larger ranks.
\par
In addition to global correlations, we also computed region-based Pearson and Spearman correlations to assess whether uncertainty reflects local reconstruction errors. Each image was divided into $100$ non\hyp overlapping patches of approximately equal size, and the correlation metrics were computed within each patch and averaged across all regions.
\par
Before computing correlation metrics, both $\tilde{q}$ and $\tilde{e}$ were blurred using a Gaussian filter with a standard deviation of $\sigma = 2$, {\color{black}chosen empirically} to suppress pixel\hyp level noise and emphasize spatially coherent patterns.

\FigFour
\FigFive

\section{Results}

\subsection{Quantitative assessment of uncertainty}

Table~\ref{tb:1} reports the mean and standard deviation of the SSIM between the reconstruction and the ground\hyp truth, as well as the mean and standard deviation of the uncertainty values estimated by our model for acceleration factors of $2\times$, $4\times$, $6\times$, $8\times$, and $10\times$ for each contrast in the brain test dataset and for acceleration factors of $4\times$ and $6\times$ for the knee test dataset. The reported uncertainty values represent the average width of the calibrated pixelwise uncertainty intervals, normalized by the maximum reconstruction magnitude, and therefore quantify the relative scale of the predicted reconstruction error as a percentage of the image intensity. For the $4\times$ accelerated reconstructions, we compared QR\hyp based and ResM\hyp based uncertainties. Both methods yielded comparable ranges for the uncertainty estimates. All uncertainty maps were normalized by the maximum reconstruction magnitude to express values as percentages. All values were computed per volumetric case.
\par
Table~\ref{tb:2} reports Pearson and Spearman correlation coefficients (including region\hyp based) between the predicted uncertainty and the true reconstruction error for the brain ($2\times$, $4\times$, $6\times$, $8\times$, and $10\times$ acceleration factors) and the knee test datasets ($4\times$ and $6\times$ acceleration factors). For the $4\times$ accelerated reconstructions, we also computed the correlation coefficients between the ResM\hyp based uncertainty and the true reconstruction error.
\par
Figure \ref{fig3:Result} shows the distribution of the Pearson correlation for all test data between the predicted uncertainty and the true reconstruction error for the $4\times$ undersampled brain (top) and knee (bottom) test dataset. Each bar represents the number of cases achieving a given correlation level. The QR\hyp based estimates (blue) are consistently higher than the ResM\hyp based ones (red) for both anatomies.

\subsection{Qualitative assessment of uncertainty}

Figure~\ref{fig4:Result} qualitatively compares the absolute reconstruction error (windowed and magnified to enhance visualization) with the QR\hyp based and ResM\hyp based uncertainty maps for a healthy and three abnormal brain cases at $4$\hyp fold undersampling. 
\par
Similarly, Figure~\ref{fig5:Result} compares the absolute reconstruction error (windowed and magnified to enhance visualization) with the QR\hyp based and ResM\hyp based uncertainty maps for three representative knee datasets at $4$\hyp fold undersampling. 
\par
Figure~\ref{fig6:Result} overlays the QR\hyp based uncertainty map on the reconstructed image for an abnormal brain and highlights the one\hyp to\hyp one correspondence between the uncertainty and pathological regions in the reconstruction. The zoomed views of the lesion provide better visualization of the subtle reconstruction differences across acceleration factors.

\subsection{Uncertainty as an anomaly detection method}

Figure~\ref{fig7:Result} presents uncertainty maps for a healthy and three representative abnormal brains. The uncertainty maps are plotted also using the maximum value from the $4\times$ accelerated cases as the lower intensity threshold. This was done to show the increase in uncertainty relative to reconstructions for acceleration factors that were considered clinically valid in 100\% of the studied cases \cite{radmanesh2022exploring}. Similarly, Figure~\ref{fig8:Result} presents thresholded uncertainty maps for three representative knees. In the bottom panel, the uncertainty maps were thresholded using the maximum value from the $4\times$ accelerated reconstructions.

\subsection{Additional Robustness Analyses}

\paragraph{Performance with Other Reconstruction Methods} {\color{black}To assess robustness across reconstruction methods, we repeated the four\hyp fold undersampled brain experiment using the FI VarNet \cite{giannakopoulos2024accelerated} instead of the E2E VarNet. We retrained the uncertainty module for FI VarNet reconstructions. The QR\hyp based uncertainty remained strongly correlated with the true reconstruction error, with Pearson, Region\hyp Pearson, Spearman, and Region\hyp Spearman coefficients of $0.91 \pm 0.10$, $0.96 \pm 0.10$, $0.75 \pm 0.18$, and $0.79 \pm 0.20$, respectively. We included a representative qualitative example in the Supporting Information \textbf{(Figure S1)}, which shows that the results are consistent with those obtained using the E2E VarNet.}

\paragraph{Out-of-Distribution Performance} {\color{black}We performed a synthetic out-of-distribution test using the $6\times$ undersampled brain setup by injecting a handcrafted bright lesion into the images. This perturbation was not present in the training data. As shown in the Supporting Information \textbf{(Figure S2)}, the QR\hyp based uncertainty increased markedly near the injected lesion, indicating sensitivity to unseen image content and highlighting the perturbed region as potentially unreliable.}

\FigSix
\FigSeven

\paragraph{Sensitivity to Sampling Pattern Variations} {\color{black}We performed a mask\hyp perturbation experiment for the $6\times$ undersampled brain setup. The uncertainty module was trained and calibrated using the Cartesian mask with 5.3\% ACS lines, and then tested on perturbed masks with 16\% and 3\% ACS lines while keeping the $6\times$ acceleration fixed. As shown in the Supporting Information \textbf{(Figure S3)}, using a mask with 5.3\% ACS lines as for training yielded the lowest uncertainty, whereas deviations from the training mask increased both the reconstruction error and the predicted uncertainty.}

\paragraph{Calibration Stability} {\color{black}We assessed the stability of $\lambda$ with respect to the size of the calibration set for the brain QR\hyp based uncertainty model at $6\times$ acceleration. We repeated the calibration procedure using one\hyp half and one\hyp quarter of the original calibration dataset (with data sampled randomly from the full dataset), which yielded $\lambda = 1.63$ and $\lambda = 1.58$, respectively. These values differ by less than $3\%$ from the value obtained with the full calibration dataset ($\lambda = 1.59$).}

\paragraph{Performance with Disjoint Training Sets} {\color{black}We performed a disjoint training experiment for the $8\times$ undersampled brain setup to evaluate whether the uncertainty module remained stable when trained on data not used to train the reconstruction network. The training and validation data were split into two non\hyp overlapping subsets using even- and odd\hyp numbered cases. E2E VarNet was trained and validated using one subset, whereas the QR\hyp based uncertainty module was trained and validated using the other subset. The experiment was then repeated with the split reversed. For this experiment we used 8 cascades for the E2E VarNet to expedite training. Both the E2E VarNet and the uncertainty module were trained using 105,000 iteration steps. A warm\hyp up ramp was applied during the first 3,750 steps, followed by cosine annealing after 70,000 steps. The rest of the training sets were kept the same as in the previous experiments. As shown in the Supporting Information \textbf{(Figure S4)}, the QR\hyp based uncertainty maps remained qualitatively similar across the two disjoint\hyp training configurations. The results suggest that the QR\hyp based uncertainty estimation is stable even when the reconstruction and uncertainty networks are trained on separate subjects.}

\section{Discussion}

Accelerated image reconstruction is one of the most transformative developments in MRI and perhaps the clearest example of an academic innovation that has successfully transitioned into widespread clinical use. Unrolled reconstruction models, such as the E2E VarNet, build directly on the principles of traditional accelerated imaging and have reduced scan times even more. Despite their success, these methods remain neural\hyp network\hyp based, and there is currently no mechanism to verify the accuracy of their outputs or detect hallucinated structures, particularly at higher acceleration factors where errors are expected \cite{radmanesh2022exploring}. In this work, we introduce an uncertainty quantification framework designed to estimate the reliability of reconstructed images for any acceleration factor. 
\par
Table~\ref{tb:1} shows that QR\hyp based uncertainty spans a similar numerical range with ResM. However, the qualitative comparison in Figures~\ref{fig4:Result} and~\ref{fig5:Result} demonstrates that QR provides uncertainty maps that more faithfully mirror the absolute reconstruction error distribution. For the healthy brain in Figure \ref{fig4:Result}, QR highlights the noisy posterior region where the reconstruction error is elevated, whereas ResM produces uniformly high uncertainty around the entire brain, inconsistent with the spatial distribution of the error. If we assume that the reconstruction error is larger in areas with pathologies, since the models were trained with cases representing only a limited number of possible brain lesions, then this suggests that the uncertainty values should be higher in correspondence of pathologies. However, we found that while both methods capture global reconstruction noise patterns, only QR consistently highlights the lesion location and structure. In the first abnormal brain, the ResM\hyp based uncertainty in the lesion is low, where the reconstruction error is large and vice\hyp versa. On the other hand, the QR\hyp based uncertainty maps have higher values at the location of the lesions and follows the spatial pattern of the reconstruction error. In the second abnormal brain, the lesion is highlighted by both methods, though its morphology is slightly distorted in ResM. In the third abnormal brain, QR captures the boundary of the treated region, which is lost in the ResM estimate. The QR\hyp based uncertainty maps also appear visually smoother than the absolute error maps because the network learns spatially coherent uncertainty patterns rather than exact pixelwise residual fluctuations. This also explains why the QR\hyp based uncertainty looks more structured than the ResM\hyp based uncertainty, which relies on direct residual magnitude regression that is more sensitive to local noise and small\hyp scale variations than quantile regression.
\par
Table~\ref{tb:2} further supports these observations by quantifying the relationship between the estimated uncertainty and the true reconstruction error calculated with fully-sampled ground truth reconstructions. After conformal calibration, the QR\hyp based uncertainty showed high Pearson correlations (roughly 90\%) with the reconstruction error in brain images for 4\hyp fold acceleration and higher, indicating strong agreement between uncertainty and error magnitude (note that the Pearson and region\hyp Pearson correlations dropped to $39\%$ and $84\%$, respectively, when computed on the unblurred maps, due to the strong influence of pixel\hyp level noise in the absolute error map). These high correlations show that the calibrated QR uncertainty scales proportionally with the underlying reconstruction error across the image. For the ResM\hyp based uncertainty, the correlation dropped to $69\%$. According to the histogram in Figure~\ref{fig3:Result}, Pearson correlation is higher when QR is used over ResM. For the reconstructed knee images, the ResM\hyp based uncertainty correlates poorly with the true reconstruction error (Pearson correlation $<25\%$) due to the smaller calibration dataset (50 volumes comparing to 689 used in the brain) \cite{angelopoulos2022image}. 
\par
Pearson correlations are consistently higher than Spearman because magnitude alignment is stronger than strict pixelwise rank ordering, which is more sensitive to local fluctuations in low\hyp error regions. The lower correlations observed at 2\hyp fold acceleration are expected, as the reconstruction error in this regime is small and spatially homogeneous, yielding insufficient dynamic range for a meaningful correlation analysis.
\par
The spatial correspondence between the lesions and regions of high uncertainty is clearly shown by overlaying the QR\hyp based uncertainty heatmap on the reconstructed images (Figure~\ref{fig6:Result}). At two\hyp fold acceleration, the uncertainty is near zero and therefore uninformative, which is expected given the high reconstruction accuracy of E2E VarNet at low acceleration factors \cite{sriram2020end}. At higher accelerations, however, the uncertainty increases, especially in the area of the pathology and in the area affected by the posterior susceptibility artifact from the prior craniotomy. Notably, the regions of elevated uncertainty precisely overlap with the true lesion locations, indicating that the QR\hyp based uncertainty behaves as a trustworthy indicator of reconstruction reliability. This alignment suggests that the QR\hyp based uncertainty can serve as a surrogate for reconstruction error during scan time, when fully sampled data are unavailable. Note that areas of elevated uncertainty are not expected to always correspond to pathological findings. If a pathology is sufficiently represented in the training data and exhibits relatively simple imaging characteristics, the reconstruction remains reliable and the corresponding uncertainty stays low, as observed for the meniscal tear example in the first row of Figure~\ref{fig5:Result}.
\par
Unrolled network architectures, such as the E2E VarNet, maintain clinical reliability for up to four\hyp fold acceleration \cite{radmanesh2022exploring}. For this reason, in Figures~\ref{fig7:Result} and ~\ref{fig8:Result}, we used the maximum uncertainty value from the four\hyp fold accelerated reconstructions as a threshold to evaluate higher\hyp acceleration factor reconstructions. In Figure ~\ref{fig7:Result}, the healthy brain example shows uncertainty below the threshold at $10\times$ acceleration, suggesting that certain slices from healthy studies could be reconstructed safely using very high acceleration factors agreeing with the findings in \cite{radmanesh2022exploring}. In the first abnormal case, uncertainty at $6\times$ undersampling remains below the threshold as well, indicating that some lesions may still be reconstructed accurately with fewer k\hyp space samples than the conventional four\hyp fold limit. In contrast, abnormal cases 2 and 3 exhibit elevated uncertainty near the lesions or in noisy regions at $6\times$ undersampling, implying that these slices should not be acquired with more than $4\times$ undersampling to ensure reliable reconstruction. {\color{black}This behavior is consistent with the objective of the QR\hyp based uncertainty module, which is trained to estimate reconstruction error intervals conditioned on the E2E VarNet output. Therefore, elevated uncertainty is expected in regions where reconstruction errors are more likely to occur, including high\hyp frequency structures, artifacts, and underrepresented pathological patterns with variable appearance. Conversely, when a pathology is well represented in the training data and reconstructed reliably, the uncertainty is not expected to increase substantially. The results of Figure~\ref{fig7:Result} suggest that highly accelerated reconstructions could potentially serve as a preliminary reliability assessment, helping identify cases or regions where a lower acceleration factor or additional k\hyp space sampling may be needed before relying on the image for diagnostic interpretation.}
\par
Conformal prediction has a few practical limitations. First, the calibration step is inherently finite\hyp sample, meaning that the estimated scaling factor can vary slightly with the size and composition of the held\hyp out calibration dataset, so small deviations from the nominal coverage level may occur in practice if the calibration data are limited or not fully representative of the test distribution. However, our robustness analysis shows that performing the calibration of the 6$\times$ brain QR\hyp based uncertainty model with reduced calibration sets changed the scaling factor by less than 3\% relative to the full set, indicating that the proposed calibration was reasonably stable with respect to calibration\hyp set size. {\color{black}Second, because the uncertainty module is conditioned on the reconstructed images, its performance is directly tied to the behavior of the reconstruction network; if the reconstruction network overfits or produces confidently wrong reconstructions, the uncertainty estimates will be unreliable.} Third, conformal prediction produces per\hyp pixel uncertainty intervals but does not directly provide a mechanism for automatically selecting or controlling the acceleration factor. Integrating uncertainty with decision rules for adaptive sampling therefore requires additional components beyond conformal prediction alone, which will be investigated in future work. Future work will also incorporate the predicted uncertainty maps into the reconstruction pipeline to selectively refine high\hyp uncertainty areas and potentially improve SSIM performance at higher accelerations.

\section{Conclusion}

\FigEight

In this work, we introduced conformal QR to quantify the uncertainty of accelerated MRI reconstruction methods in the absence of ground\hyp truth references. Our results demonstrate that QR provides a faithful surrogate for reconstruction error, with strong agreement in both magnitude and spatial distribution across contrasts and acceleration factors. QR\hyp based uncertainty consistently highlights pathological regions and reconstruction failures that are missed by simpler heuristic notions of uncertainty. Since uncertainty increases with undersampling and is higher in regions with pathologies, it could be combined with highly accelerated scans for rapid anomaly detection. More broadly, real\hyp time access to uncertainty information could enable adaptive sampling strategies in which the scanner acquires additional k\hyp space lines only when needed, thereby optimizing scan efficiency while preserving diagnostic quality. {\color{black}To facilitate reproducibility and future comparisons with alternative uncertainty quantification methods, including posterior\hyp sampling\hyp based approaches, we are making our code and execution scripts used in this study openly available.}

\section*{Acknowledgments}
The authors thank Dr. Tsui-Wei (Lily) Weng for valuable discussions.

\section*{Data Availability Statement}
   The PyTorch code for our models is available at \href{https://github.com/IliasGiannakopoulosLab/VarNet}{https://github.com/IliasGiannakopoulosLab/VarNet}.

\Urlmuskip=0mu plus 1mu\relax
\bibliographystyle{IEEEtran}


\end{document}